\documentclass{aa}
\usepackage{graphicx}
\begin{document}
    \title{Quasi-hydrostatic intracluster
gas under radiative cooling}
    \subtitle{}

    \author{K. Masai
           \inst{1}
           \and
           T. Kitayama\inst{2}
           }

    \offprints{K. Masai}

    \institute{Department of Physics, Tokyo Metropolitan University,
               Hachioji, Tokyo 192-0397, Japan\\
               \email{masai@phys.metro-u.ac.jp}
          \and
              Department of Physics, Toho University,
              Miyama, Funabashi, Chiba 274-8510, Japan\\
              \email{kitayama@ph.sci.toho-u.ac.jp}
              }

    \date{Received , 2003; accepted , 2004}

    \abstract{Quasi-hydrostatic cooling of the intracluster gas is
studied.  In the quasi-hydrostatic model, work done by gravity on the
inflow gas with $dP \neq 0$, where $P$ is
the gas pressure, is taken into account
in the thermal balance.  The gas flows in
from the outer part so as to compensate the pressure loss of
the gas undergoing radiative cooling, but the mass flow is so 
moderate
and smooth that the gas is considered to be quasi-hydrostatic.
The temperature of the cooling gas decreases toward the cluster center,
but, unlike cooling flows with $dP = 0$, approaches a constant temperature of 
$\sim 1/3$ the temperature of the non-cooling
ambient gas.  This does not mean that gravitational work cancels out radiative
cooling, but means that the temperature of the cooling gas 
appears to approach a constant 
value toward the cluster center if the gas maintains the quasi-hydrostatic balance.
We discuss the mass flow in quasi-hydrostatic cooling,
and compare it with the standard isobaric cooling flow model.
We also discuss the implication of $\dot{M}$ for the standard cooling
flow model.

    }

    \maketitle

\section{Introduction}

The intracluster gas, i.e. hot gas in clusters of galaxies, which undergoes
radiative cooling is thought to flow in toward the cluster center to
maintain pressure equilibrium (see Fabian \cite{Fabian94} for a
review and references therein).  According to the standard cooling flow
picture, the mass deposition rate $\dot{M}$ can be estimated from the
differential luminosity.  X-ray observations showed that $\dot{M}$
inferred from the surface brightness profile was over 
100~$M_{\sun}$~yr$^{-1}$
for so called cooling flow clusters.  However,
firm evidence has not been found for such a large amount of cooled gas that
should be detected at longer wavelengths if it existed in the cluster
core (e.g., Edge
\cite{Edge01}).

{\it ASCA} and recent {\it Chandra} and {\it XMM-Newton} observations 
(e.g.,
David et al. \cite{David01}; Kaastra et al. \cite{Kaastra01};
Allen et al. \cite{Allen01}; Peterson et al. \cite{Peterson03})
with better spatial and spectral resolution reveal, however, that
$\dot{M}$ is smaller by
an order of magnitude than what was estimated
before from the surface brightness.
Moreover, spectroscopic analysis shows that the gas
temperature decreases toward the cluster center, as expected for
cooling flows, but down to only half or one-third the temperature of 
the non-cooling ambient gas.  This implies the
lack of soft X-ray emission, or no appreciable contribution of
lower temperature  gas.

To explain the observations, various
theoretical models have been proposed (see e.g., B\"{o}hringer et al.
\cite{Boehringer02}; Peterson et al. \cite{Peterson03}).  Many of those
consider the central active galaxy (with an active galactic nucleus (AGN)) as a 
heating source to compensate radiation loss (e.g., Churazov et al.
\cite{Churazov01}; Ruszkowski \& Begelman \cite{Ruszkowski02}; Kaiser \&
Binney \cite{Kaiser03}).  This idea is interesting from the aspect of
history of activity (Reynolds, Heinz \& Begelman \cite{Reynolds02}),
since many cooling flow clusters possess cD galaxies in their central
regions.  However, some fine tuning is required.  
While the radiative cooling rate is determined by the 
local values of temperature and density, the conductive or convective 
heating rate including hot bubbles from AGN depends on the scale length 
of the temperature or pressure as well.  
This could make the processes complicated, and the parameters
such as diffusion coefficient or mixing length remain open (Narayan \&
Medvedev \cite{Narayan01}; Voigt et al. \cite{Voigt02}).

In a standard analysis of X-ray data, an isobaric cooling flow model is
applied to estimate $\dot{M}$ (Fabian \cite{Fabian94}).
However, analysis based on surface brightness and analysis based on spectrum
give inconsistent results, as mentioned above, in spite of both being
based on the same physical concept.
This may suggest some
problem underlying the standard cooling flow model.
In the standard isobaric model, work done by gravity on the inflow gas 
is not properly taken into account.  This is not an additional heating source
but is naturally expected for the gas under gravity.  If the inflow is
induced and regulated smoothly by local radiative cooling under gravity,
quasi-hydrostatic structure may be attained.  Such a simple picture is
thought to be a starting point for considering an alternative heat source.

In the present paper we examine quasi-hydrostatic
intracluster gas undergoing radiative cooling,
taking proper account of the work done by gravity.  In Sect. 2 we present
our quasi-hydrostatic cooling model. 
We compare our model with the standard isobaric cooling flow model in
Sect. 3, and give concluding remarks in Sect. 4.

\section{Quasi-hydrostatic cooling model}

The internal energy $U$ of the gas changes as
\begin{equation}
dU = C_V dT = d'Q - Pd{1 \over \rho}
\label{PdV}
\end{equation}
where $C_V$ is the specific heat at constant volume, and $T$, $\rho$ 
and $P$
are the gas temperature, density and pressure, respectively.  The
first and second terms on the right-hand side represent heat flowing into
and work done on the gas, respectively.  If the gas is
flowing toward the cluster center, work by gravitational force would
be done on the gas.

We consider a spherically symmetric structure.  By using the EOS
(equation of state) of an ideal gas and the hydrostatic balance equation
\begin{equation}
{1 \over \rho}dP = -{GM_r^* \over r^2}dr,
\label{dP/rho}
\end{equation}
Eq. (\ref{PdV}) is rewritten for the quasi-hydrostatic intracluster
gas, as
\begin{equation}
{5 \over 2}k{M \over \mu m} dT = -Ldt - {GM_r^* \over r^2} M dr.
\label{Ldt+Fdr}
\end{equation}
Here $M(r)$ is a gas mass undergoing radiation loss at a
rate $L(r)$; one can consider, for instance, a shell
as $M(r)=4\pi r^2\rho{\mit\Delta}r$.
$M_r^*$ is the gravitational mass
contained in the radius concerned,
$\mu$ the mean molecular weight, $m$ the proton mass, and other
quantities have their usual meanings.
The term `quasi-hydrostatic' does not mean static but is used for the 
steady
state under gravity.

From Eq. (\ref{Ldt+Fdr}) and the
continuity equation,
\begin{equation}
\dot{M}_r = -4\pi r^2 \rho {dr \over dt},
\label{continuity}
\end{equation}
we obtain
\begin{equation}
\dot{M}_r \left({5 \over 2}{k \over \mu m}{dT \over dr}+{GM_r^*
\over r^2}\right) = 4\pi r^{2}\rho{L \over M} = 4\pi r^2n^2\Lambda,
\label{Mdot}
\end{equation}
where $n = \rho/\mu m$ is the number density and $n^2\Lambda$ the 
cooling
rate per unit volume.  $\dot{M}_r$ represents the gas mass flowing
from the outer part into $r$ per unit time.

We assume that the intracluster gas is
isothermal before cooling when virialized, i.e., $(3/2)(kT/\mu m) \sim -\phi$ for the 
gravitational
potential $\phi$ in the cluster (see e.g., Sarazin \cite{Sarazin86}).
In such a case, the gravitational mass
$M_r^*$, which includes the galaxies and the dark matter as well as the 
gas,
can be replaced with the equilibrium (virial) temperature $T_0$, as
\begin{equation}
{GM_r^* \over r} \simeq {3 \over 2} \sigma_r^2
= {3 \over 2} \beta {kT_0 \over \mu m},
\label{T0}
\end{equation}
where $\sigma_r$ is the line-of-sight velocity dispersion of the galaxies
in the cluster; $\beta$ is a parameter in the so-called 
$\beta$-model of
intracluster gas.
We consider that $M_r^*$ is large enough compared to the gas mass $M_r$
within $r$
and that the gas inflow hardly alters the gravitational potential.

Eq. (\ref{Mdot}) with $T_0$ is expressed in the form
\begin{eqnarray}
{d\ln T \over d\ln r} & = & {9 \over 5} \left[{4\pi \over 3}r^{3}
{n^{2}\Lambda \over (3/2)kT}{\mu m \over \dot{M}_r} -
{1 \over 3} \beta {T_0 \over T}
\right] \nonumber \\
& = & {9 \over 5} \left[{\tilde{M}_r/\tilde{t}_{cool}
\over \dot{M}_r} - {1 \over 3} \beta {T_0 \over T}\right],
\label{dlnT/dlnr}
\end{eqnarray}
where, in the last expression,
\begin{equation}
\tilde{M}_r \equiv {4\pi \over 3}r^3\rho \quad\mbox{and}\quad
\tilde{t}_{cool} \equiv {(3/2)nkT \over n^{2}\Lambda}.
\label{t_cool}
\end{equation}
Hereafter the quantities with tildes represent those evaluated with
the {\em local} values of density and temperature;
$\tilde{M}_r$ and $\tilde{t}_{cool}$ are the mass of a uniform gas 
sphere
and the cooling time, which are given by $\rho(r)$ and $T(r)$.
Accordingly, $\tilde{M}_r$ represents the gas mass
within $r$ that cools radiatively
in the cooling time $\tilde{t}_{cool}$ given at $r$.
Eq. (\ref{Mdot}) or (\ref{dlnT/dlnr}) shows the relation between variations
$dT$ and $dr$ to satisfy Eq. (\ref{PdV}) for the relation between 
$\dot{M}_r$ and $L$ under quasi-hydrostatic balance,
i.e., hydrostatic balance (Eq. (\ref{dP/rho})) with 
a mass flow (Eq. (\ref{continuity})).

We consider that the gas inflow across $r$ is
caused by radiative cooling within $r$
under quasi-hydrostatic balance at every $r$.
While the gas of mass $\tilde{M}_r$ inside loses its pressure
on a time scale $(\partial\ln P/\partial t)^{-1} \sim
(\partial\ln T/\partial t)^{-1} = \tilde{t}_{cool}$,
gas of temperature $T$ comes into the region at rate $\dot{M}_r$ 
to compensate the loss with the time scale $\tilde{t}_{cool}$. 
The inflow mass at $r$ during $\tilde{t}_{cool}$ is expressed
approximately as
\begin{equation}
\dot{M}_r \tilde{t}_{cool}  \sim
{d\tilde{M}_r \over dt} \tilde{t}_{cool} \sim \tilde{M}_r.
\end{equation}
Therefore, we assume $\dot{M}_r \sim \tilde{M}_r/\tilde{t}_{cool}$
for a local quasi-static change at a given radius.
This condition for observable quantities and 
its validity are discussed in detail later.
It should be noted that $\tilde{M}_r$ differs from the gas mass $M_r$
contained within $r$, as it depends on the density profile $\rho(r)$.
The cooling time of the gas $M_r$ is also different from $\tilde{t}_{cool}$.
We mention later such global quantities in the discussion of the above condition.

The gas that undergoes radiative cooling has a temperature
profile given by Eq. (\ref{dlnT/dlnr}), while the ambient gas that is
free from radiative cooling remains isothermal.
We define a radius $r_{cool}$ at which $\tilde{t}_{cool} \sim t_H =
H_0^{-1}$, the Hubble time.  $\tilde{t}_{cool} < t_H$ at $r < 
r_{cool}$,
and $\tilde{t}_{cool} > t_H$ at $r > r_{cool}$, because 
$\tilde{t}_{cool}$
becomes monotonically shorter toward the inner region.
At $r > r_{cool} \sim$~100 kpc, radiative
cooling has little effect on the hydrostatic structure, and the gas
temperature is nearly equal to the gravitational temperature.  On the
other hand, at
$r < r_{cool}$ the effect of cooling becomes important.
The temperature decreases and the gas is flowing inward, 
yet the gas is likely quasi-hydrostatic if the inflow is smooth and its
speed is much slower than the local sound speed $\propto \sqrt{T}$, or
more correctly, the ram pressure is small compared to the
thermal pressure.

Thus, the temperature of the quasi-hydrostatic cooling gas
begins to decrease from $T_0$ at $r_{cool}$,
and approaches an asymptotic value given by
\begin{equation}
T \sim {1 \over 3} \beta T_0 {\dot{M}_r \over 
\tilde{M}_r/\tilde{t}_{cool}}
\label{minT}
\end{equation}
toward the cluster center.
\begin{figure}[htb]
\includegraphics[scale=.78]{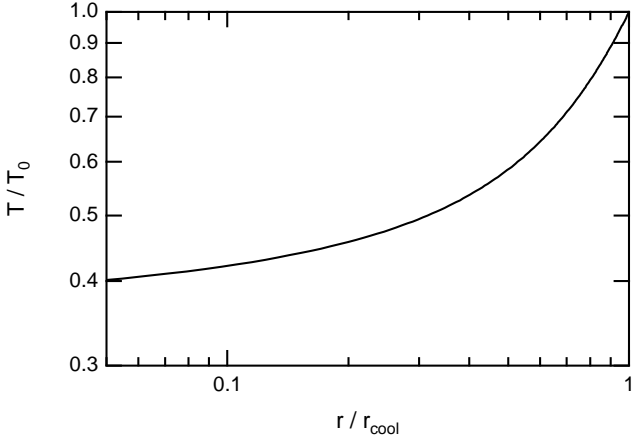}
\caption{The temperature of cooling quasi-hydrostatic gas, as a function
of $r/r_{cool}$ for the case
$\dot{M}_r=\beta^{-1}\tilde{M}_r/\tilde{t}_{cool}$.  This corresponds,
for instance, to $\dot{M}_r=\tilde{M}_r/\tilde{t}_{cool}$ for $\beta=1$ 
and
to $\dot{M}_r=1.5\tilde{M}_r/\tilde{t}_{cool}$ for $\beta=2/3$.}
\end{figure}
In Fig.~1 we show a temperature profile given by Eq. (\ref{dlnT/dlnr})
with $\dot{M}_r \sim \tilde{M}_r/\tilde{t}_{cool}$ at every $r$.
For $\dot{M}_r = \beta^{-1}\tilde{M}_r/\tilde{t}_{cool}$,
the temperature decreases as $T \propto r^{6/5}$ at $T \sim T_0$, as $T
\propto r^{3/5}$ at $T \sim (1/2) T_0$, and eventually becomes $\sim
(1/3) T_0$.  In practice, at the cluster center, the gas accumulates
and cools further to a lower temperature.  It is likely that at $T 
\la T_0$ ($r
\la r_{cool}$) the temperature is somewhat less steep than $T
\propto r^{6/5}$ and becomes flatter outward so that the temperature
connects smoothly with $T \sim T_0$ at $r \sim r_{cool}$.
If electron conduction works, the temperature would vary more gradually
at $0.5 \la r/r_{cool} \la 1$.

We examine the density profile of the gas
in quasi-hydrostatic
cooling considered here.  For simplicity, we express the density and
temperature profiles as $\rho \propto r^\alpha$ and $T \propto r^\eta$,
where $\eta$ is given by Eq. (\ref{dlnT/dlnr}).  Since
\begin{equation}
\tilde{M}_r \propto r^{3+\alpha} \quad\mbox{and}\quad
\tilde{t}_{cool} \propto r^{-\alpha+\eta(1-\xi)}
\end{equation}
with $\Lambda \propto T^\xi$, we have
\begin{eqnarray}
{\tilde{M}_r/\tilde{t}_{cool} \over \dot{M}_r} \propto
r^{1+\alpha-\eta(1-\xi)-\zeta}
\end{eqnarray}
with the inflow velocity $\propto r^\zeta$.
Here $1-\xi > 0$ in the temperature range of interest
and it is likely $\zeta < 0$ for smooth inflow.
For a quasi-static change at given $r$, that is, the condition
$\dot{M}_r \sim \tilde{M}_r/\tilde{t}_{cool}$ means
\begin{equation}
1+\alpha-\eta(1-\xi)-\zeta \sim 0
\end{equation}
with a small value of $|\zeta|$ ($= r/L_v)$ compared to unity, where
$L_v$ is the scale length of the inflow velocity.

Thus, when the gas is cooling with quasi-hydrostatic balancing,
the density varies as
\begin{equation}
\rho \propto r^{-1+\eta(1-\xi)+\zeta},
\end{equation}
roughly as $\rho \propto r^{-0.5}$ at $r \la r_{cool}$
and approaches $\propto r^{-1}$ for
$\eta \to 0$ toward the cluster center.
Note that
the local cooling time is independent of the temperature/density 
profile,
\begin{equation}
\tilde{t}_{cool} \propto r^{1-\zeta},
\end{equation}
and varies roughly as $\tilde{t}_{cool} \propto r$.
Inversely, if observations show $\tilde{t}_{cool} \propto r$,
this implies $-\alpha+\eta(1-\xi) \sim 1$ and therefore
the validity of the assumption $\dot{M} \sim 
\tilde{M}_r/\tilde{t}_{cool}$
at every $r$ or $(\tilde{M}_r/\tilde{t}_{cool})/\dot{M} \sim$ constant.
A recent report on 16 clusters observed by {\it Chandra}
(Voigt \& Fabian \cite{Voigt04})
seems to suggest $\tilde{t}_{cool} \propto r$.
These properties as well as the temperature profile predicted by
the quasi-hydrostatic cooling model with
$\dot{M}_r \sim \tilde{M}_r/\tilde{t}_{cool}$
can be confirmed by observations.
The gas mass $M_r$ contained within $r$ is larger than $\tilde{M}_r$
for the density profile with $\alpha < 0$.
With $t_{cool}$ being the global cooling time of the gas $M_r$, as
\begin{eqnarray}
&& M_r = \int 4\pi r^2 \rho dr \sim {3 \over 3+\alpha} \tilde{M}_r
\nonumber \\ \mbox{and} \nonumber \\
&& t_{cool} = {\int 4\pi r^2 (3/2)nkT dr \over \int 4\pi r^2 n^2\Lambda 
dr}
\sim {3+2\alpha+\eta\xi \over 3+\alpha+\eta} \tilde{t}_{cool},
\nonumber
\end{eqnarray}
the difference between $M_r/t_{cool}$ and 
$\tilde{M}_r/\tilde{t}_{cool}$ is
estimated to be within a factor of $\sim 2$.

While the gas is cooling, $\dot{M}_r$ is coming in 
from the outer part so as to maintain the quasi-hydrostatic balance.
Such a flow is not very drastic, since the gas is cooling monotonically
and thereby the inflow is smooth.  Thus, the mass flow
rate follows
\begin{equation}
\dot{M}_r = {6 \over 5}\tilde{L}_r{\mu m \over kT} \left({d\ln T \over
d\ln r}+{3 \over 5} \beta {T_0 \over T}\right)^{-1},
\label{MdotLT0}
\end{equation}
where $\tilde{L}_r \equiv (4\pi/3)r^3n^2\Lambda$ is given
by the density and temperature at $r$.
$\dot{M}_r$ is in the range
(2/3--2)$\cdot\tilde{L}_r(\mu m/kT_0)$ for $T_0$
and does not vary very much
through the flow (see Sect. 4).
For $\dot{M}_r \sim \tilde{M}_r/\tilde{t}_{cool}$, Eq.
(\ref{MdotLT0}) reduces to
\begin{eqnarray}
\dot{M}_r & \sim & {2 \over 3}\tilde{L}_r{\mu m \over kT}
\label{MdotL} \\
& \sim & 8.4 \left({r_{cool} \over \mbox{100
kpc}}\right)^3 \left({n \over 10^{-3} \mbox{ cm$^{-3}$}}\right)^2
\left({kT_0 \over \mbox{keV}}\right)^{-1} \nonumber \\
& \quad & \times \left({\Lambda \over 10^{-23} \mbox{ erg s$^{-1}$
cm$^3$}}\right) \mbox{$M_{\sun}$ yr$^{-1}$},
\end{eqnarray}
where $\mu = 0.61$ is assumed.

For the values of typical cooling flow clusters, the mass
flow rate may be $\sim 100 M_{\sun}$~yr$^{-1}$.
The local inflow rate $\dot{M}_r$
does not mean the mass deposition rate (see Sect. 3)
but merely gives an upper limit on the cooled mass in the cluster.
Eq. (\ref{MdotLT0}) or (\ref{MdotL}) means that 
following the change in $T$ (Eqs. (\ref{Ldt+Fdr}) and (\ref{Mdot})) 
$\dot{M}_r$ increases with increasing $\tilde{L}_r$.
This is reasonable if the
mass inflow is induced by radiation loss; $\dot{M}_r$ is not simply
proportional to $\tilde{L}_r$ but depends also on $T$.  The dependence 
in Eq.
(\ref{MdotL}) should be compared with that
in the standard isobaric cooling flow model, which is
discussed in the following section.

\section{Comparison with isobaric cooling flow model}

In the quasi-hydrostatic model, work done by gravity on the
intracluster gas is taken into account.  The pressure is
given by Eq. (\ref{dP/rho}), and therefore $dP \neq 0$ in this model.
In comparison with quasi-hydrostatic cooling, we discuss isobaric
cooling which is often applied to estimate the mass flow rate from
X-ray observations.

For given mass $M$, with $dP = 0$ Eq. (\ref{PdV}) gives
\begin{equation}
{5 \over 2}k{M \over \mu m} dT = d'Q = -Ldt.
\label{Ldt}
\end{equation}
This is nothing but the definition of the specific heat $C_P = 
(d'Q/dT)_P$
at constant pressure.  Eq. (\ref{Ldt}) simply expresses that for
mass $M$ the temperature decreases with time due to radiation loss
under $dP = 0$.  The density must increase to maintain the pressure
equilibrium, and the cooling time becomes shorter as $\propto
T^2/\Lambda$.  Thus, the gas cools and flows in increasingly toward the
cluster center.

Eq. (\ref{Ldt}) may be compared with the expression
for the standard cooling flow model (Fabian \cite{Fabian94}),
\begin{eqnarray}
{5 \over 2}k{\dot{M}_{CF} \over \mu m} dT = dL, \nonumber
\end{eqnarray}
where $dT$ is defined for a change from
$T$ to $T-dT$ (Johnstone et al. \cite{Johnstone92}; Fabian
\cite{Fabian94}), i.e. $dT > 0$ for cooling, and corresponds
to the variation $-dT$ in the present paper.
So that, instead of the above equation, we consider
\begin{equation}
{5 \over 2}k{\dot{M}_{CF} \over \mu m} dT = -dL
\label{-dL}
\end{equation}
as the standard cooling flow model.  With increasing radiation
loss ($dL > 0$), the temperature decreases ($dT < 0$) for given
$\dot{M}_{CF}$ ($> 0$); otherwise, if $\dot{M}_{CF} < 0$
were the case, the temperature would increase for $dL > 0$.

According to the standard cooling flow model (Fabian \cite{Fabian94}),
$dL$ in Eq. (\ref{-dL}) is the differential luminosity defined
as $dL = (dL/dT)dT$.
Therefore, Eq. (\ref{-dL}) is not a derivative of Eq. (\ref{Ldt}).
This implies that Eq. (\ref{-dL}) comes from some
other concept than $dP = 0$ in Eq. (\ref{PdV}).
$\dot{M}_{CF}$ may be interpreted
as the gas mass that is cooling and
dropping out of the flow per unit time (see Wise \& Sarazin
\cite{Wise93}) or the mass deposition rate
(Johnstone et al. \cite{Johnstone92}).

For the enthalpy of the cooling gas of given $T$,
with $dP=0$ we may have
\begin{equation}
{5 \over 2}k{\dot{M} \over \mu m} T \sim -L,
\label{enthalpy}
\end{equation}
consistently with Eq. (\ref{Ldt}) for given $M$.
Eq. (\ref{enthalpy}) means $\dot{M} < 0$ 
(cf. $\dot{M}_{CF} > 0$) or
that the gas of temperature $T$ decreases at a rate $\propto L$.
The lost gas must be so cool, much cooler than $T$, 
that  it no longer contributes to $L$ at all.

Suppose that the intracluster gas has temperature $T$ to emit $L$.
With $M_d$ being the cooled mass, it increases at a rate
roughly as $\dot{M}_d \sim - \dot{M} > 0$ in the cluster.
So that, for the mass deposition rate $\dot{M}_d$,
we have a global relation
\begin{eqnarray}
{5 \over 2}k{\dot{M}_d \over \mu m} T \sim L,
\label{MdotL2}
\end{eqnarray}
where the left-hand side does not mean the enthalpy of the cooling gas; 
remember that the gas of $M_d$ should be too cool ($\ll T$) to be responsible for $L$.
Even if $\dot{M}_{CF}$ is
considered to be $\dot{M}_d$,
however, it is unlikely that Eq. (\ref{MdotL2})
reduces to Eq. (\ref{-dL})
for thermodynamical variation $dT $.
As readily seen, Eq. (\ref{MdotL2})
gives $\dot{M}_d \propto L/T$, while $\dot{M}_{CF} \propto dL/dT$
in Eq. (\ref{-dL}).
Accordingly, the standard cooling flow model would yield an $\dot{M}_{CF}$
that traces $d\Lambda/dT$, which is determined by atomic processes
regardless of the thermal properties of the intracluster gas,
as discussed below.

Eq. (\ref{-dL}) means that
$\dot{M}_{CF}$ is proportional directly to $-dL/dT$, because
$dL$ is the variation with $T$ in the standard cooling flow model.
Considering $L \propto \rho^2 \Lambda (M/\rho) \propto \rho\Lambda M$
with $P = \mbox{const.} \equiv P_0$,
we can express the luminosity in the form
\begin{eqnarray}
L = {M \over \mu m}{\Lambda \over kT} P_0. \nonumber
\end{eqnarray}
Thus, Eq. (\ref{-dL}) can be rewritten as
\begin{eqnarray}
{5 \over 2}k{\dot{M}_{CF} \over \mu m} & = & -{dL \over dT}
\label{-dL/dT} \\
& = & {L \over T}\left(1-{d\ln \Lambda \over d\ln T}\right) \\
& \sim  & {5 \over 2}k{\dot{M}_d \over \mu m}
\left(1-{d\ln \Lambda \over d\ln T}\right),
\end{eqnarray}
where $1-d\ln\Lambda/d\ln T > 0$ in the temperature range of interest.
With decreasing
$T$, the sign of $d\ln \Lambda/d\ln T$ changes to be negative at $kT
\sim 2$ keV, below which line cooling dominates over bremsstrahlung.
While $\dot{M}_{CF} \sim 0.5 \dot{M}_d$ at $kT \ga 2$ keV,
$\dot{M}_{CF} \sim 1.5 \dot{M}_d$ at $kT \la 2$ keV,
on account of the factor $(1-d\ln\Lambda/d\ln T)$.
Consequently, $\dot{M}_{CF}$ would 
vary sensitively to local minima/maxima in $\Lambda(T)$.

Eq. (\ref{-dL/dT}) has been applied to estimate the mass deposition rate
from the soft X-ray luminosity (e.g. Peterson et al.
\cite{Peterson03}).
In a limited photon-energy range, the $T$ dependence of $\Lambda$ becomes
complicated because of the presence of lines.  Therefore, observations
with different energy bands could derive different values of
$\dot{M}_{CF}$, or spectral analyses could derive values different from
each other.  In any case, $\dot{M}_{CF}$ of the standard cooling flow
model seems not to give a reasonable representation of the mass
deposition rate due to radiation loss $d'Q = -Ldt$.
If the luminosity $L$ of a cluster is well accounted for by emission of 
the intracluster gas of nearly isothermal temperature $T$, Eq. 
(\ref{MdotL2})
would give roughly the isobaric deposition rate of the cooled-down mass
in the cluster.

\section{Discussion and remarks}

In the quasi-hydrostatic model, we consider the ideal condition that a
change is quasi-static.  The pressure loss due to
radiative cooling is immediately compensated by continuous, smooth,
spherical mass inflow from the outer part.  Neither turbulence nor
overshooting occurs during the process.  The inflow is so moderate as
not to disturb the hydrostatic balance significantly.  Thus, if the
local inflow rate is perfectly controlled by the local cooling rate, the
gas is quasi-hydrostatic and $\dot{M}_r \sim 
\tilde{M}_r/\tilde{t}_{cool}$
is attained; 
we will investigate this point further by means of hydrodynamical
calculations elsewhere.

As mentioned in Sect. 2, `quasi-hydrostatic' means not
static but steady-state.  Accordingly, although the temperature of
the cooling gas approaches a constant value
toward the cluster center (Fig. 1), this does not mean
that gravitational work can cancel out radiative cooling.  This
temperature profile is no more than a result of the steady inflow to
maintain the hydrostatic balance.
It should be noted that the gravitational potential or $M^*_r$ is
assumed to be little affected by the gas inflow.

At the cluster center (or close to
a cD galaxy) where the inflow mass is accumulating, the cooling
time can be too short for the actual flow rate to follow the
cooling rate of the gas inside, i.e.
$\tilde{M}_r/\tilde{t}_{cool} > \dot{M}_r$.
Then the quasi-hydrostatic balance would break, and the gas temperature
inside decreases below $\sim (1/3)\beta T_0$, yet the temperature 
in the
steady flow region will continue to follow Eq. (\ref{dlnT/dlnr}).

In the opposite case, if the cooling rate at a given radius is
smaller than the inflow rate as 
$\tilde{M}_r/\tilde{t}_{cool} < 
\dot{M}_r$, the flow rate $\dot{M}_r$
would decrease inward and the temperature
approaches a constant value higher than $\sim (1/3)\beta T_0$ toward
the cluster center.  This may be the case when cooling in the core is
not yet very significant or when some heating works against local
radiative cooling so that $\tilde{t}_{cool}$ is effectively longer
than given by Eq. (\ref{t_cool}).

The gas gains momentum during inflow.  In the quasi-static
model this is included into Eq. (\ref{dP/rho}), which
determines the hydrostatic structure in cooperation with EOS.  This is
valid as long as the momentum flux is small compared to the thermal
pressure.  In practice, however, particularly in the region where the
temperature approaches a constant value (see Fig. 1), the inflow gas
may partly overshoot because of a fluctuation about $\dot{M}_r \sim
\tilde{M}_r/\tilde{t}_{cool}$.  This may lead to instability and
result in a break in
the spherically symmetric structure.  The resultant dense part cools
more rapidly, but the gas remains
quasi-hydrostatic on the whole if the momentum flux is still small on 
average.

Within the context of quasi-hydrostatic cooling, the gas inflow must be 
mild
at every radius.
However, if the flow evolved so that the momentum flux would
become considerable, though still subsonic,
the quasi-hydrostatic balance would fail unless some force
works against the inertia.  In such a case, for the
gas to be quasi-hydrostatic, some momentum flux 
not in the form of the thermal pressure by heat may be needed against
the inflow.

Observations by coming missions can examine the radial profiles
predicted by the quasi-hydrostatic
cooling model for several quantities, such as
local cooling time $\tilde{t}_{cool}$, mass flow rate
$\dot{M}_r$ or local luminosity $\tilde{L}_r$,
and possibly inflow velocity as well as the density and temperature.
The observed temperature profiles can be understood invoking 
quasi-hydrostatic cooling, though  some momentum/heat
source may be working additionally.  The simple quasi-hydrostatic
model is expected to give a basis for understanding the thermal
properties of the intracluster gas under gravity and for advanced study
using hydrodynamics codes.

\begin{acknowledgements}
The authors would like to thank Naomi Ota for discussion about the 
structure of intracluster gas and her helpful comments.
Part of this work was supported by the
Grant in Aid (14740133, 15037206) for Scientific Research of the
Ministry of Education and Science in Japan.
\end{acknowledgements}


\begin{thebibliography}{}

\bibitem[2001]{Allen01} Allen S.W., Schmidt R.W., Fabian A.C., 2001,
MNRAS, 328, L37

\bibitem[2002]{Boehringer02} B\"{o}hringer, H.,
Matsushita, K., Churazov, E., Ikebe, Y., Chen, Y., A\&A, 382, 804

\bibitem[2001]{Churazov01} Churazov, E., Sunyaev, R., Forman, W.,
B\"{o}hringer, H., 2001, ApJ, 554, 261

\bibitem[2001]{David01} David, L. P., Nulsen, P. E. J., McNamara, B.
R., Forman, W., Jones, C., Ponman, T., Robertson, B., Wise, M., 2001,
ApJ, 557, 546

\bibitem[2001]{Edge01} Edge, A. C., 2001, MNRAS, 328, 762

\bibitem[1994]{Fabian94} Fabian, A. C. 1994, ARA\&A, 32, 227

\bibitem[1992]{Johnstone92} Johnstone, R. M., Fabian, A. C., Edge, A.
C., Thomas, P. A., 1992, MNRAS, 255, 431

\bibitem[2001]{Kaastra01} Kaastra, J. S.,
Ferrigno, C., Tamura, T., Paerels, F. B. S., Peterson, J. R., Mittaz,
J. P. D., A\&A, 365, L99

\bibitem[2003]{Kaiser03} Kaiser, C., Binney, J., 2003, MNRAS, 338, 837

\bibitem[2001]{Narayan01} Narayan R.,
Medvedev, M. V., 2001, ApJ, 562, L129

\bibitem[2003]{Peterson03} Peterson, J. R., Kahn,
S. M., Paerels, F. B., Kaastra, J. S., Tamura, T., Bleeker, A. M.,
Ferrigno, C., \& Jernigan, J. G. 2003, ApJ, 590, 207

\bibitem[2002]{Reynolds02} Reynolds, C. R., Heinz, S., Begelman, M.
C., 2002, MNRAS, 332, 271

\bibitem[2002]{Ruszkowski02} Ruszkowski, M., Begelman, M. C., 2002,
ApJ, 581, 223

\bibitem[1986]{Sarazin86} Sarazin, C. L., 1986, Rev. Mod. Phys., 58, 1

\bibitem[2002]{Voigt02} Voigt, L. M., Schmidt, R. W., Fabian, A. C.,
Allen, S. W., Johnstone, R. M., 2002, MNRAS, 335, L7

\bibitem[2004]{Voigt04} Voigt, L. M., Fabian, A. C., 2004, MNRAS, 347, 
1130

\bibitem[1993]{Wise93} Wise, M. W., Sarazin, C. L., 1993,
ApJ, 415, 58

\end{thebibliography}
\end{document}